\documentclass[aps,pre,twocolumn,showpacs]{revtex4}
\usepackage{amsmath,amsfonts,bm, color,amssymb}
\usepackage{graphicx,epsfig,overpic,hyperref,ulem}
\usepackage{color}

\newcommand{\refeq}[1]{Eq. \ref{#1}}
\newcommand{\sigm}{{\sigma}}
\newcommand{\eps}{{\varepsilon}}

\input epsf 

\newcommand{\Rr}{\mbox{\it R}}
\newcommand{\Sr}{\mbox{\it S}}
\newcommand{\out}{{\mathrm{ex}}}
\newcommand{\ins}{{\mathrm{in}}}

\newcommand{\bE}{{\bf E}}
\newcommand{\bP}{{\bf{P}}}

\newcommand{\eq}{{\mathrm{eq}}}

\begin{document}

\title{Electrohydrodynamic Quincke rotation of  a  prolate ellipsoid}
\author{ Quentin Brosseau, Gregory Hickey, Petia M. Vlahovska}
\affiliation{%
 School of Engineering, Brown University, Providence, RI 02912, USA
}

\date{\today}

\begin{abstract}
We experimentally study the occurrence of spontaneous spinning (Quincke rotation)  of an ellipsoid in a uniform DC electric field. 
 For an ellipsoid suspended  in an unbounded fluid,  we find two stable states characterized by the orientation of the ellipsoid long axis relative to the applied electric field : spinless (parallel) and spinning (perpendicular). The phase diagram of ellipsoid behavior as a function of field strength and aspect ratio is in close agreement with the theory of Cebers et al. Phys. Rev .E 63:016301 (2000).   We also investigated the dynamics of the ellipsoidal Quincke  rotor resting on a planar surface with normal perpendicular to  the field direction. We find novel behaviors, such as swinging (long axis oscillating around the applied field direction) and tumbling,  due to the confinement.

\end{abstract}

\maketitle

\section{Introduction}

The spontaneous spinning of a dielectric sphere in a uniform DC electric field has been known for over a century, first attributed to the work of Quincke in 1896 \cite{Quincke:1896}.
This phenomenon is enjoying resurgent interest driven by the experimentally observed intriguing dynamics of Quincke-rotating spheres  \cite{Lemaire:2002, Lemaire:2005, Jakli:2008, Bartolo:2013} and  drops \cite{Sato:2006, Salipante-Vlahovska:2010, Salipante-Vlahovska:2013, Ouriemi:2014,Ouriemi:2015},
 as well as  an increasing interest in electric field directed assembly of colloidal particles \cite{Velev_review:2015, Blaaderen:2013, Aranson:2013, Dobnikar:2013}. Theoretical analyses of rigid spheres show that an isolated sphere can exhibit Lorenz chaotic rotations \cite{Lemaire:2005}, pairs of spheres can undergo intricate trajectories \cite{Das-Saintillan:2013, Dolinsky-Elperin:2012,Lushi-Vlahovska:2014},  large populations can self-organize and undergo directed motion \cite{Bartolo:2013, Bartolo:2015, Belovs:2014, Yeo-Lushi-Vlahovska:2014}, and a suspension can exhibit lower effective viscosity \cite{Cebers:2004, Lemaire:2008, Huang-Zahn-Lemaire:2011}  or increased conductivity \cite{Lemaire:2009b} compared to the suspending fluid.

The Quincke  rotation arises from particle polarization in an applied electric field due to the accumulation of free charges at the particle interface  \cite{Melcher-Taylor:1969}.
The induced dipole due to these free charges lags the application of the uniform DC electric field. The evolution of the $i$-th dipole component in a coordinate system centered at the particle (and in the case of an ellipsoid aligned with its axes, see Figure \ref{figE}) is described by \cite{Cebers:2000}
\begin{equation}
\label{Pevol}
 \begin{split}
\frac{d P_i}{d t}&=-\tau_i^{-1}\left(P_i-\left(\chi^0_i-\chi^\infty_i\right)E_i\right)\,,
\end{split}
\end{equation}
where  $\chi^0_i,\, \chi^\infty_i$ are the low- and high-frequency susceptibilities and $\tau_i$ the Maxwell-Wagner polarization time along the $i$-th axis of the ellipsoid.

For a shape-isotropic particle such as a sphere  with radius $a$ 
\begin{equation}
\label{Peff2}
\begin{split}
\chi_0-\chi_\infty= 4 \pi \eps_\out a^3  \frac{3(\Rr-\Sr)}{(\Rr+2)(\Sr+2)}\,,
\end{split}
\end{equation}
and the Maxwell-Wagner polarization time is
 \begin{equation}
\label{tMW}
\tau=\frac{\eps_{\ins}+2\eps_{\out}}{\sigm_{\ins}+2\sigm_{\out}}\,.
\end{equation}
The subscripts $``\ins "$ and $``\out "$ denote the values for particle and suspending medium, respectively.  $\Rr$ and $\Sr$ characterize the mismatch of  electrical conductivity, $\sigm$,  and permittivity, $\eps$ 
\begin{equation}
\Rr=\frac{\sigm_\ins}{\sigm_\out}\,,\quad \Sr=\frac{\eps_\ins}{\eps_\out}\,.
\label{ParameterRatios}
\end{equation}
 Analysis of the polarization relaxation equations,  \refeq{Pevol},  together with the balance of electric and viscous torques shows that a spinning state is  possible above a critical field strength, which for a sphere is given by  \cite{Jones:1984, Turcu:1987}
\begin{equation}
\label{EQ}
E_Q^2=-\frac{\alpha}{\tau\left(\chi_0-\chi_\infty\right) }=\frac{2\sigm_\out \mu_\out \left(\Rr+2\right)^2}{3\eps_\out^2 (\Sr-\Rr )}\,,
\end{equation}
\refeq{EQ} shows that the electrorotation is possible only if $\left(\chi_0-\chi_\infty\right) <0$, i.e., the suspending fluid and sphere satisfy the condition $\Rr/\Sr<1$. Physically, this corresponds to the induced dipole moment of the sphere oriented opposite to the direction of the applied field. A perturbation in the dipole orientation produces  torque, which induces physical rotation of the sphere (around an axis perpendicular to the applied field direction).  The induced surface-charge distribution rotates with the sphere, however,  the exterior fluid  recharges the interface.  The balance between charge convection by rotation and supply by conduction from the bulk results in a steady oblique orientation of the dipole relative to the field direction and constant torque.

\begin{figure}[h!]
\centerline{\includegraphics[width=3.5in]{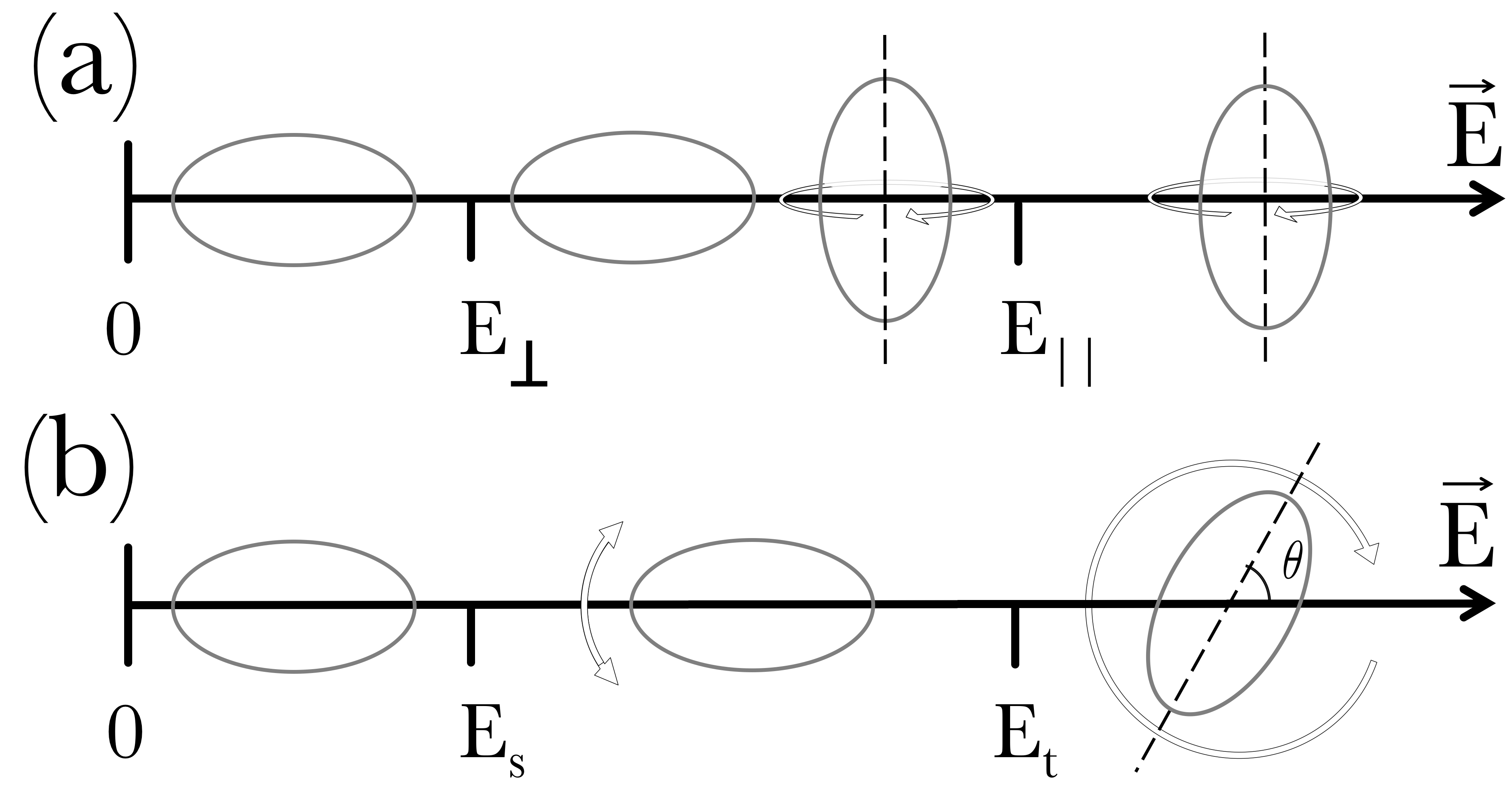}}
\caption{\footnotesize Summary of a prolate ellipsoid stable steady-states and threshold electric fields. (a) Bulk: spinless-parallel, coexistence of spinless-parallel and spinning-perpendicular: Spinning-perpendicular.  (b) Resting on a surface:  spinless-parallel, swinging and tumbling.}
\label{States}
\end{figure}

An ellipsoidal particle is expected to display more complex electrorotation dynamics than a sphere \cite{Cebers:2000, Cebers:2002, Dolinsky-Elperin:2005, Dolinsky-Elperin:2006,Dolinsky-Elperin:2009b, Dolinsky-Elperin:2009} due to its anisotropic shape. The polarization relaxation times and susceptibilities along the long and short axes differ.
The classic result is that an axisymmetric ellipsoid aligns its long axis with the applied electric field \cite{JonesTB, Cebers:2000}. 
In stronger fields, $E>E_\perp$, a stable orientation perpendicular to the external DC field is predicted \cite{Cebers:2000}
\begin{equation}
\label{Ec1}
E^2_{\perp}=-\frac{\alpha_{\parallel}}{\tau_{\perp}\left(\chi^0_\perp-\chi^\infty_\perp\right)}\,, 
\end{equation}
where $``i=\perp''$ denotes the ellipsoid  symmetry  axis and $``i=\perp"$ - the perpendicular-to-symmetry axis. $\alpha_{i}$ 
is the friction coefficients for rotation around the $i-$th axis \refeq{frics_coefs}.
 In this state, the  prolate ellipsoid is rotating  around its long axis (spinning state), see Figure \ref{States}.a for an illustration.
For a certain range of electric field strengths the system is bistable, i.e.,  the two orientations--parallel and transversal to the applied field-- coexist. 
The bistability disappears above $E>E_{||}$,
\begin{equation}
\label{Ec2}
E^2_{\parallel}=\frac{\alpha_{\perp}}{\alpha_{\parallel}}E^2_{\perp}\frac{\chi^\infty_\perp-\chi^0_\perp}{\chi^\infty_\perp-\chi^0_{||}}\,.
\end{equation}

 At even higher field strengths, time integration of the equations of motion and polarization relaxation \cite{Adamthesis} shows 
that  the ellipsoid axis can begin to precess around the field direction; regimes of both regular and chaotic oscillations are found with increasing field strength \cite{Cebers:1991, Cebers:1993, Adamthesis}.  
    
These theoretically predicted peculiar features of the electrorotation  of ellipsoids have not been experimentally verified. In this paper, we perform a systematic study of ellipsoid dynamics as a function of field strength and aspect ratio for prolate ellipsoid.

\section{Experimental methods}

\subsection{Preparation of the ellipsoid}

The millimeter sized ellipsoid, needed for direct visualization, is made by  cross-linking a spheroidally deformed drop of  polymer NOA 81 (Norland Optical Adhesive). The drop deformation is induced by application of a uniform DC electric field after the drop is suspended in  PDMS 500 cSt (UTC) in a set-up similar to \cite{Salipante-Vlahovska:2010, Ouriemi:2014}. Once steady-state deformation is reached, the drop is exposed to UV radiation $365nm$ for about a $1\, min$. Aspect ratios ranging from 1 to 2 can be achieved by tuning the electric field strength, from 0 to 5.0 $kV/m$.
 Once cross-linked, the solid NOA particles ($\eps_{NOA}=4.04\eps_0$,  $\sigm_{NOA}=10^{-17}$ S/m, and  density $\rho_{NOA}\sim1.2 g/cm^3$)  are rinsed and transferred to the Castor oil bath ($\eps_{CO}=4.8\eps_0$, $\sigm_{CO}=4.4\times 10^{-11}$S/m, $\rho_{CO}=0.961\, g/cm^3$, viscosity $\mu_{CO}=0.69$Pa.s), where the experiments are performed. 

\begin{figure}[h!]
\centerline{\includegraphics[width=2.5in]{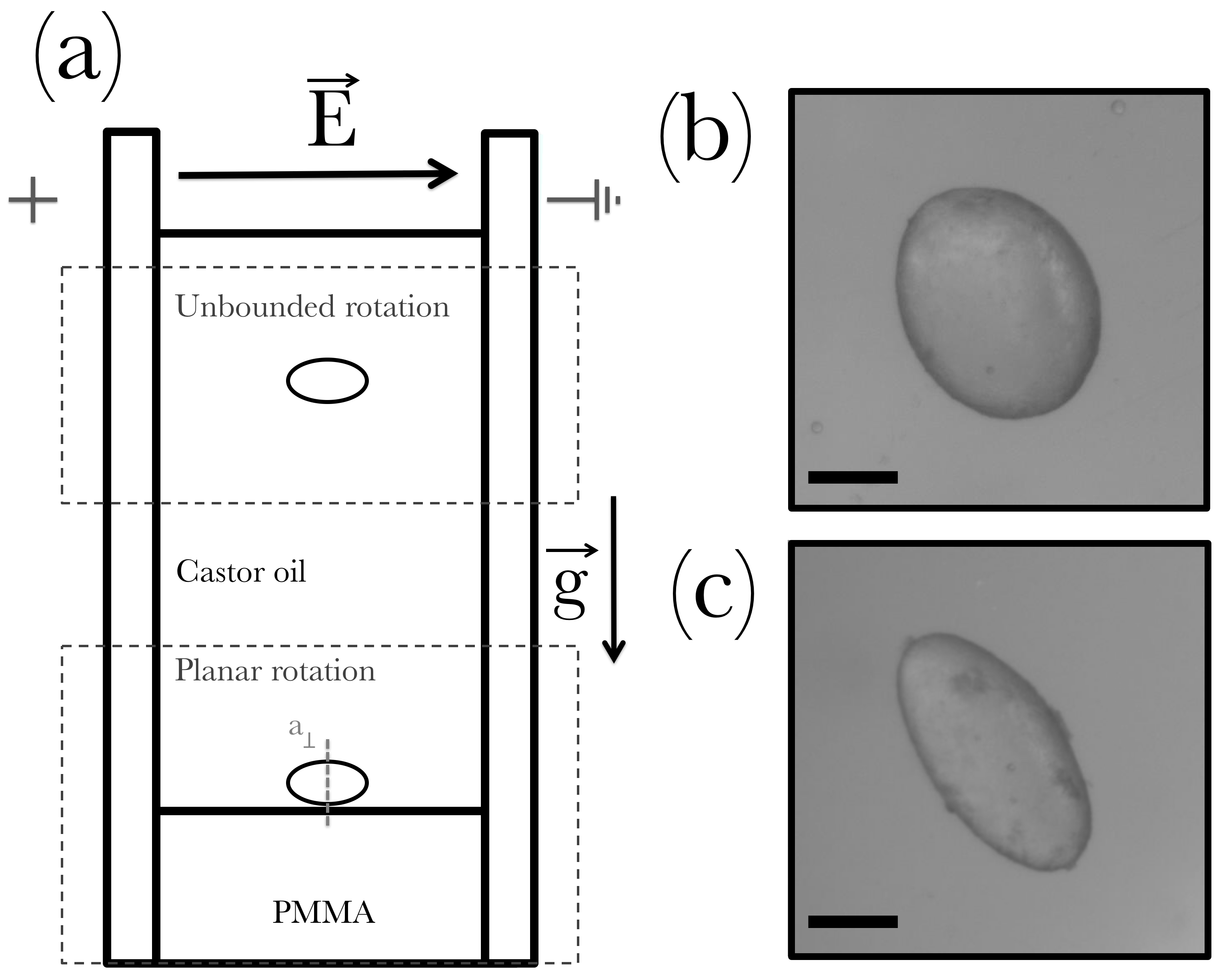}}
\caption{\footnotesize Sketch of the experimental set-up. Ellipsoid  rotation is observed either in the bulk 
or while resting  on a horizontal PMMA surface. Top view of two NOA ellipsoids with aspect ratio $\beta=1.25$ (b), $\beta=1.93$ (c). Scale bar is $500\mu m$.}
\label{setup}
\end{figure}

\subsection{Experimental set up and procedure}

A uniform electric field is generated in a parallel-plate chamber set-up sketched in Figure \ref{setup}.a.  
The chamber base and walls are constructed from polymethyl methacrylate (PMMA).
Two $4\times 7$ cm brass plates  serving as electrodes are attached to the chamber vertical walls.  The  distance between the electrodes is 1.64 $cm$, about 10 times larger than the particle radius, to minimize boundary effects. Fields up to 3.0 $kV/cm$ are generated using a voltage amplifier connected to a DC power supply. The ellipsoid is observed from the top, in a direction perpendicular to the field.

The experiment is carried out by placing the solid NOA ellipsoid into the chamber filled with castor oil. The particle is either placed in the bulk (to observe unbounded dynamics) or at the bottom (to observe planar rotations). After the electric field is applied, the ellipsoid behavior is recorded for about 3 minutes.  
Due to slight density mismatch the particle sediments in the bulk with a very slow speed (about $0.4mm/s$).

Motion of the particle is recorded through an optical device mounted on top of the chamber. The frames obtained are then treated numerically in order to extract the parameters of interests that are: particle geometry (volume and aspect ratio) and the dynamic angle $\theta (t)$ between the particle symmetry axis $a_{\parallel}$ and the applied electric field  $E$.

\section{Results and discussion}

\subsection{Unbounded rotation}
\begin{figure}[h!]
\centerline{\includegraphics[width=2.7in]{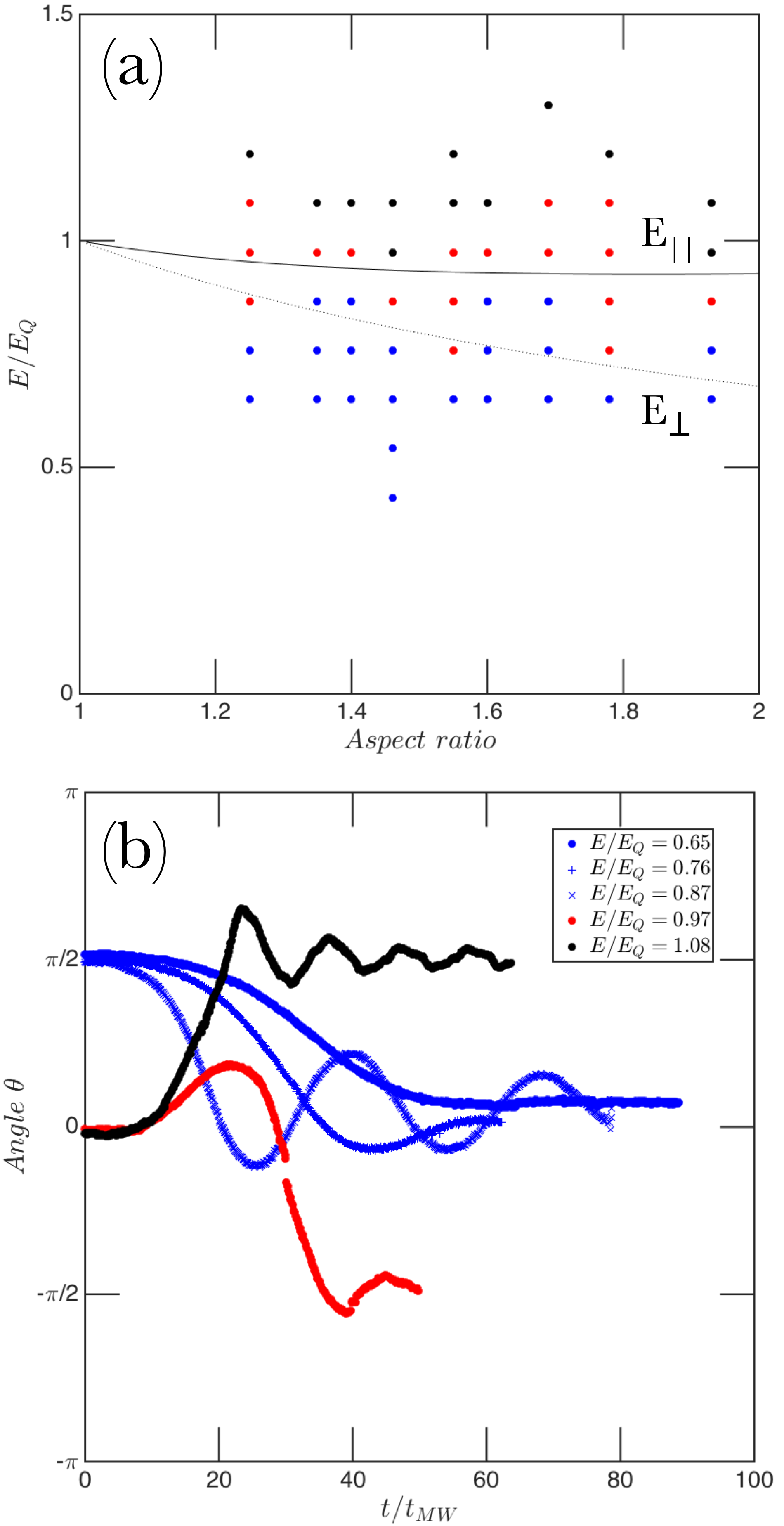}}
\caption{\footnotesize (a) Phase diagram of unbounded dynamics of a prolate ellipsoid showing the two steady states: spinless-parallel orientation (blue) and spinning-perpendicular state (red and black). Curves corresponds to stability thresholds of spinning-perpendicular state $E_{\perp }$(doted \refeq{Ec1}), and the loss of stability of the spinless-parallel state $E_{\parallel}$ (dashed \refeq{Ec2}) (b) The 
relaxation dynamics depends on the magnitude of the electric field and it is illustrated for a particle with aspect ratio $\beta=1.40$.}
\label{3D}
\end{figure}

Figure \ref{3D}.a shows the steady-states for ellipsoids in unbounded environment  as a function of field strength.  9  ellipsoids with  aspect ratios between   $\beta=1.02$  and $\beta=1.96$ are investigated.  We find the spinless-parallel state (blue), and the spinning-perpendicular state (red and black). The spinning-perpendicular state appears   as soon as the threshold for its stability $E_{\perp}$ (doted line), \refeq{Ec1},  is exceeded. The spinless-parallel state is never observed above $E_{\parallel}$ (dashed line), \refeq{Ec2}, its predicted stability limit \cite{Cebers:2000}. As a result of $E_{\parallel }>E_{\perp }$, both rotation and alignment of identical ellipsoids can be observed in the bistability region (supplementary material). The even distribution of the events shows that sedimentation of the particle favors no specific state. 

Our experiments highlight the rich transient dynamics as the particle approaches the steady state. The various relaxations are shown on Figure\ref{3D}.b for an ellipsoid  with aspect ratio $\beta=1.40$. 

If the applied field is $E<E_{\perp }$ , the particle is initially placed with its long axis perpendicular to the field direction ($\theta(t=0)=\pi/2$).  
When the field is turned on, the steady spineless state is reached monotonically at low field strengths. In stronger fields, the steady orientation 
is approached via damped oscillations around the parallel orientation.

If the applied field is  $E>E_{\perp }$, the particle is initially placed in parallel alignment ($\theta(t=0)=0$) and its relaxation  towards the spinning-perpendicular orientation is recorded. 
Here again two  possible relaxation behaviors exist.  
Slightly above the threshold value $E_{\parallel }$ (Figure \ref{3D}.b red curve), the long axis exhibits growing oscillations until  the spinning motion stabilizes the orientation to $\theta=\pi/2$ \cite{Cebers:2002}. If the particle initial orientation is $\theta(t=0)=\pi/2$ then spinning takes place immediately and this initial orientation is maintained (see movie in the Supplementary material).
In the second case, decaying oscillations  around $\theta=\pi/2$ are observed ( black curve) which correspond to out-of-plane precession of the symmetry axis.

\subsection{Planar rotations}

\begin{figure}[h!]
\centerline{\includegraphics[width=2.7in]{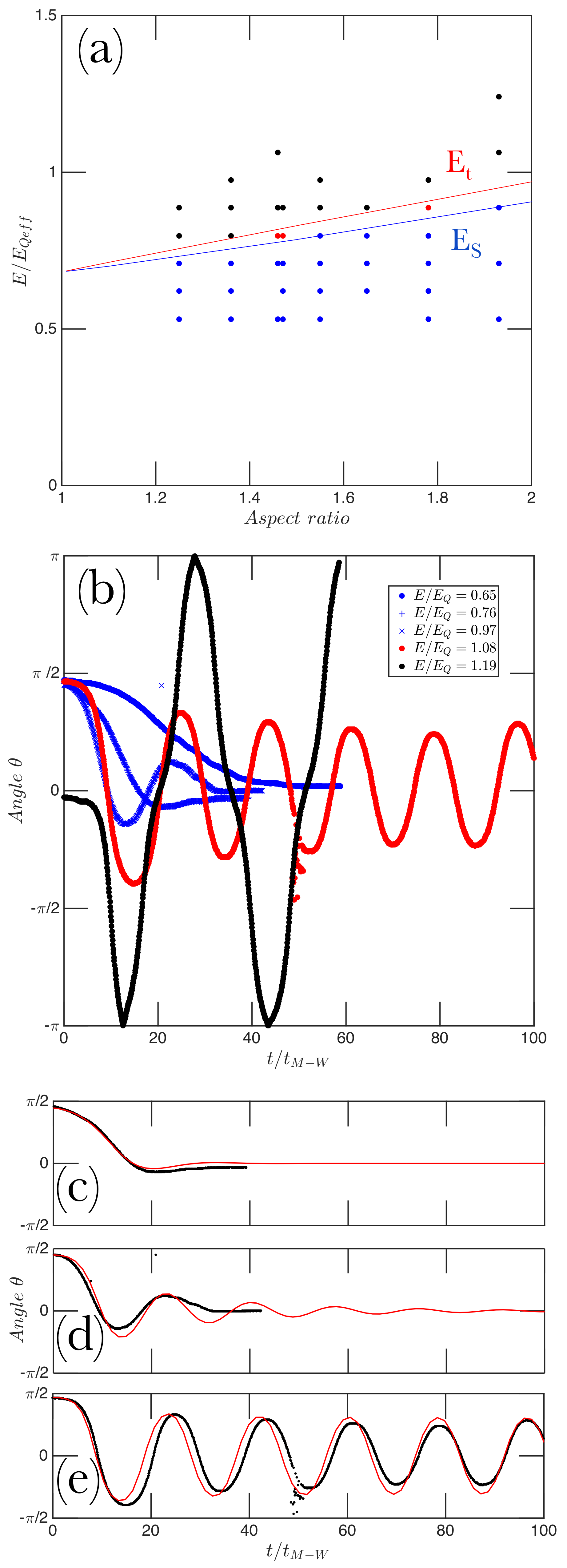}}
\caption{\footnotesize (a) Phase diagram for a planar rotation: spinless-parallel (blue), swinging (red) and tumbling (black). Numerically obtained thresholds for swinging (red) and tumbling (black). (b) Example of transient dynamics for ellipsoid with aspect ratio $\beta=1.78$. Swinging (here $-\pi/4<\theta<\pi/4$)occurs for $E/E_Q=1.08$ and the tumbling for $E/E_Q=1.19$. The best numerical agreements are found for an effective conductivity of the surrounding material $\epsilon_{eff}=0.67\epsilon_{CO}$; for $E/E_Q=0.76$ (c), $E/E_Q=0.97$ (d) and $E/E_Q=1.08$ (d).}
\label{2D}
\end{figure}

In this Section we consider  the rotation a prolate ellipsoid  resting on a planar surface, see Figure \ref{setup}.a,  in order to study rotations restricted to a plane (i.e., suppressing the transition to the spinning-perpendicular state). Under these conditions, numerical simulations show that the particle dynamics has three distinct stable states: spinless-parallel, swinging  and tumbling, see Figure \ref{States}.b.

Experiments confirm this scenario. The phase diagram  in Figure \ref{2D}.a shows the regions of existence of these three states as the field strength increases: a spinless-parallel state (blue), sustained swinging motion (red) , and   tumbling. Unlike unbounded rotation the distinct states are not observed to coexist. 

The transient dynamics is presented in Figure \ref{2D}.b for a particle of aspect ratio $\beta=1.78$. Relaxation to spinless-parallel state also exhibit non-monotonous dynamics for fields larger than $E/E_Q=0.76$ (blue). Swinging and tumbling states occurs independently of the initial orientation of the particle. Swinging (here between $-\pi/4<\theta<\pi/4$) sustains a constant amplitude as long as the electric field is on. The amplitude can be tuned by modifying the strength of the electric field and the particle aspect ratio. 
Tumbling (black), occurs when amplitude of swinging goes beyond $\pi/2$ and can take both rotation directions. However near the threshold  $E_{t}$ the ellipsoid might reverse its direction after a few cycles. 

The experimental data is compared to  the theoretical model, which is summarized in Appendix A. 
A quantitative agreement for thresholds (Figure \ref{2D}.a)  and relaxations (Figure \ref{2D}.c.d.e) is  found if an effective conductivity of the outer fluid $\epsilon_{eff}=0.67\epsilon_{CO}$ is assumed. This assumption is justified by  the fact that the  substrate introduces inhomogeneity in the conductivity  near the ellipsoid: a substrate with a negligible  conductivity can be accounted for by  an effectively lower  conductivity.  
Some discrepancies may also originate from the additional friction due to roughness on particles surface and substrate. The overall agreement proves that numerical model of a one axis rotation is sufficient to describe this configuration. 

Thresholds $E_{s}$ and $E_{t}$ are found experimentally to be lower that $E_Q$.  This is in agreement with the behavior of ellipsoidally deformed drops \cite{Salipante-Vlahovska:2010}

\section{Conclusions}

We have investigated the electrorotation dynamics of a prolate ellipsoid under unbounded or constrained (planar) conditions. As predicted by stability analysis ellipsoid has the ability to either align or spin perpendicularly to the electric field. Spinning-perpendicular states dominates at high electric fields and a bistability region was evidenced. Restricting the rotations from fully 3D to planar, by placing the ellipsoid on a surface give rise to new dynamics such as tumbling and swinging. 

This work focused prolate ellipsoids, where rotation around the longest axis is favored due to the lower friction. Oblate ellipsoids offer potentially reacher behavior with choice of ``rolling" (spinning around the shortest, symmetry axis) or ``tumbling'' (spinning around the longer axis).

The electrorotation dynamics   of ellipsoids resting on a surface offers interesting directions for research in the context of collective dynamics. The shape anisotropy adds aligning interactions, which lacks in systems of Quincke rollers \cite{Bartolo:2013}, and this could give rise to nematic ordering.

\section{Acknowledgments}
This work was supported by NSF-CBET Awards   1437545 and 1544196. 

\appendix
\section{Quincke rotation of an ellipsoid}

\begin{figure}[h!]
\centerline{\includegraphics[width=2.0in]{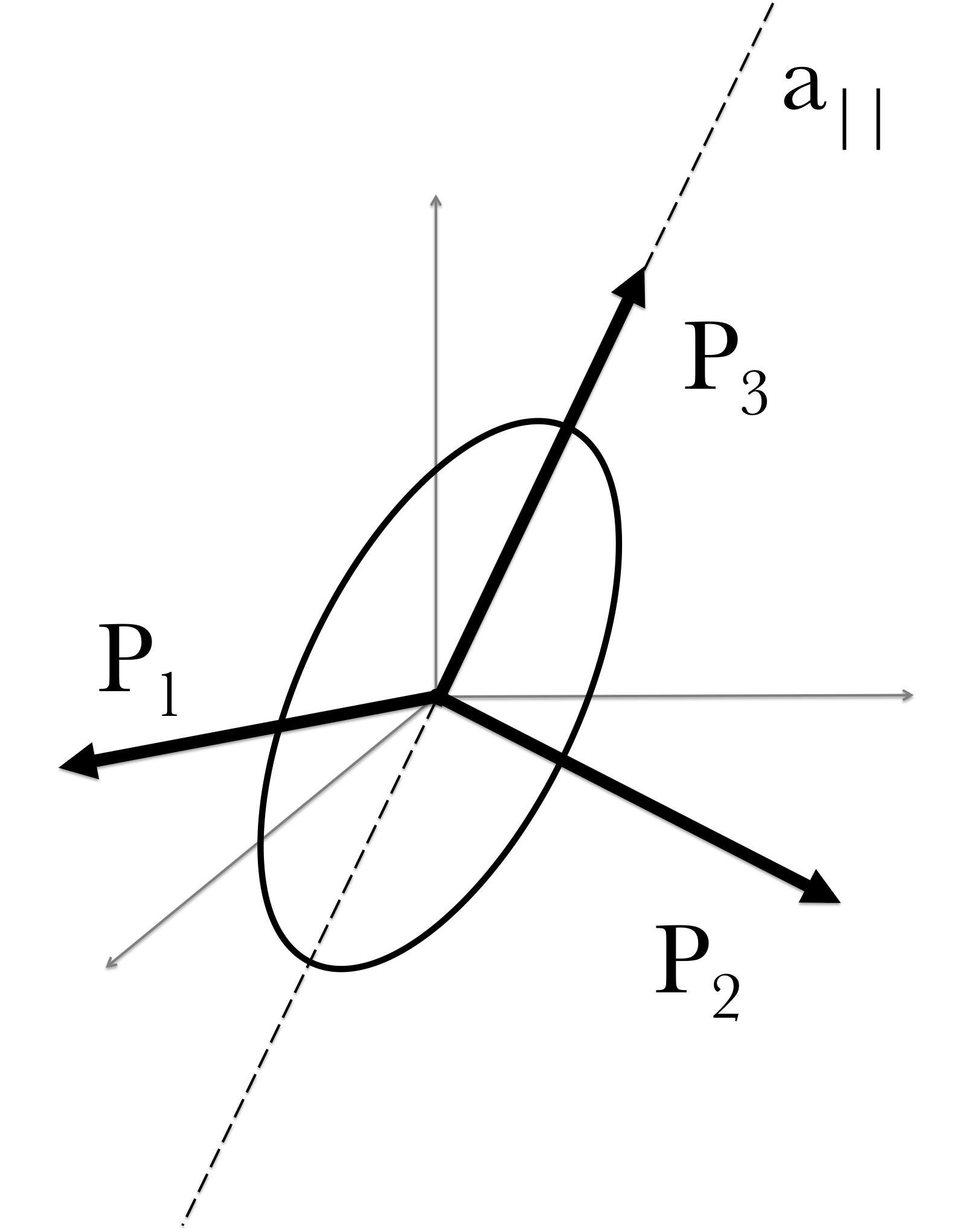}}
\caption{\footnotesize Reference frame of the ellipsoid, cartesian appears in grey and corrotating frame $P_{i}$ in black. }
\label{figE}
\end{figure}

Here we summarize  the model for ellipsoid \cite{Cebers:2002}. 
We consider a coordinate system aligned with the ellipsoid symmetry axis (and co-moving with it).
Charges carried by conduction  accumulate at boundaries that separate media with different electric properties.  The   relaxation  of the retarded polarization (total minus the instantaneous) along the direction of the ellipsoid axes s  described  by
\begin{equation}
\label{Prel}
\frac{d P_i}{d t}=-\frac{1}{\tau_i}\left(P_i-P_i^{\eq}\right)
\end{equation}
where
\begin{equation}
\tau_i=\frac{\eps_\out}{\sigm_\out}\frac{\left(2+\Sr\right)\left(1+n_i\left(\Rr-1\right)\right)}{\left(2+\Rr\right)\left(1+n_i\left(\Sr-1\right)\right)}
\end{equation}
are the Maxwell-Wagner relaxation times and $n_i$ are the depolarization factors
$n_i$. $P^{\eq}_i$ are the components of the equilibrium polarization, which are proportional to the components of the electric field. The coefficient of proportionality is the susceptibility
\begin{equation}
P^{eq}_i=\chi_i E_i\,,
\end{equation}
\begin{equation}
\label{chi_def}
 \chi_i=\chi^0-\chi^\infty=\eps_\out V \frac{(2+\Rr)(2+\Sr)}{9\left[1+n_i\left(\Rr-1\right)\right]\left[1+n_i\left(\Sr-1\right)\right]}
\end{equation}

For a {\it{prolate}} ellipsoid $a_3>a_2=a_1$ (i.e., $a_{||}>a_{\perp}$), the axis of symmetry is the long axis; the aspect ratio $\beta=a_{||}/a_\perp>1$
\begin{equation}
n_3(\equiv n_{||})=\frac{1-e^2}{e^3}\left[\mbox{arctanh}(e)-e\right]\,,\quad n_1=n_2=(1-n_3)/2
\end{equation}
where $e=\sqrt{1-1/\beta^2}$

 The evolution of the ellipsoid orientation is determined from the 
 conservation of angular momentum 
\begin{equation}
\label{angular conserv}
I_i\frac{d\Omega_i}{d t}=\left(\bP_t\times\bE\right)_i- 2\mu_\out V \alpha_i\Omega_i\,.
\end{equation}
where the  friction coefficients
\begin{equation}
\alpha_1=\frac{1+\beta^2}{n_{||} \beta^+\frac{1-n_{||}}{2}}\,,\quad \alpha_2=\alpha_1\,,\quad \alpha_3=\frac{2}{1-n_{||}}
\label{frics_coefs}
\end{equation}
For planar  (2D) rotations,  $\Omega_1=d \theta/dt$, $\Omega_2=\Omega_3=0$, $E_1=0$, $E_2=-E \sin\theta$ and $E_3=E\cos\theta$

\bibliographystyle{unsrt}

\begin{thebibliography}{10}

\bibitem{Quincke:1896}
Quincke G.
\newblock Ueber rotation em im constanten electrischen felde.
\newblock {\em Ann. Phys. Chem.}, 59:417--86, 1896.

\bibitem{Lemaire:2002}
E.~Lemaire and L.~Lobry.
\newblock Chaotic behavior in electro-rotation.
\newblock {\em Physica A}, 314(1-4):663--671, November 2002.

\bibitem{Lemaire:2005}
F.~Peters, L.~Lobry, and E.~Lemaire.
\newblock Experimental observation of lorenz chaos in the quincke rotor
  dynamics.
\newblock {\em Chaos}, 15:013102, 2005.

\bibitem{Jakli:2008}
A.~Jakli, B.~Senyuk, G.X. Liao, and O.~Lavrentovich.
\newblock Colloidal micromotor in smectic a liquid crystal driven by dc
  electric field.
\newblock {\em Soft Matter}, 4:2471--2474, 2008.

\bibitem{Bartolo:2013}
A.~Bricard, J.-B. Caussin, N.~Desreumaux, O.~Dauchot, and D.~Bartolo.
\newblock {Emergence of macroscopic directed motion in populations of motile
  colloids}.
\newblock {\em {Nature}}, {503}({7474}):{95--98}, {Nov 7} {2013}.

\bibitem{Sato:2006}
H.~Sato, N.~Kaji, T.~Mochizuki, and Y.~H. Mori.
\newblock Behavior of oblately deformed droplets in an immiscible dielectric
  liquid under a steady and uniform electric field.
\newblock {\em Phys. Fluids}, 18:127101, 2006.

\bibitem{Salipante-Vlahovska:2010}
P.~F. Salipante and P.~M. Vlahovska.
\newblock Electrohydrodynamics of drops in strong uniform dc electric fields.
\newblock {\em Phys. Fluids}, 22:112110, 2010.

\bibitem{Salipante-Vlahovska:2013}
P.~F. Salipante and P.~M. Vlahovska.
\newblock Electrohydrodynamic rotations of a viscous droplet.
\newblock {\em Phys. Rev. E}, 88:043003, 2013.

\bibitem{Ouriemi:2014}
M.~Ouriemi and P.~M. Vlahovska.
\newblock Electrohydrodynamics of particle-covered drops.
\newblock {\em J. Fluid Mech.}, 751:106--120, 2014.

\bibitem{Ouriemi:2015}
M.~Ouriemi and P.~M. Vlahovska.
\newblock Electrohydrodynamic deformation and rotation of a particle-coated
  drop.
\newblock {\em Langmuir}, 31:6298--6305, 2015.

\bibitem{Velev_review:2015}
Bhuvnesh Bharti and Orlin~D. Velev.
\newblock {Assembly of Reconfigurable Colloidal Structures by Multidirectional
  Field-Induced Interactions}.
\newblock {\em {Langmuir}}, {31}({29}):{7897--7908}, {JUL 28} {2015}.

\bibitem{Blaaderen:2013}
A.~van Blaaderen, M.~Dijkstra, R.~van Roij, A.~Imhof, M.~Kamp, B.~W. Kwaadgras,
  T.~Vissers, and B.~Liu.
\newblock {Manipulating the self assembly of colloids in electric fields}.
\newblock {\em {Eur. Phys. J -Special Topics}}, {222}({11}):{2895--2909}, {NOV}
  {2013}.

\bibitem{Aranson:2013}
Igor~S. Aranson.
\newblock {Collective behavior in out-of-equilibrium colloidal suspensions}.
\newblock {\em {C. R. Physique}}, {14}({6}):{518--527}, {JUN-JUL} {2013}.

\bibitem{Dobnikar:2013}
Jure Dobnikar, Alexey Snezhko, and Anand Yethiraj.
\newblock {Emergent colloidal dynamics in electromagnetic fields}.
\newblock {\em {Soft Matter}}, {9}({14}):{3693--3704}, {2013}.

\bibitem{Das-Saintillan:2013}
D.~Das and D.~Saintillan.
\newblock {Electrohydrodynamic interaction of spherical particles under Quincke
  rotation}.
\newblock {\em {Phys. Rev. E}}, {87}({4}), {APR 29} {2013}.

\bibitem{Dolinsky-Elperin:2012}
Yu. Dolinsky and T.~Elperin.
\newblock {Dipole interaction of the Quincke rotating particles}.
\newblock {\em {PHYSICAL REVIEW E}}, {85}({2, 2}), {FEB 27} {2012}.

\bibitem{Lushi-Vlahovska:2014}
E.~Lushi and P.~M. Vlahovska.
\newblock Periodic and chaotic orbits of micro-rotors in creeping flows.
\newblock {\em Journal of Nonlinear Science}, 25:1111--1123, 2015.

\bibitem{Bartolo:2015}
Antoine Bricard, Jean-Baptiste Caussin, Debasish Das, Charles Savoie,
  Vijayakumar Chikkadi, Kyohei Shitara, Oleksandr Chepizhko, Fernando Peruani,
  David Saintillan, and Denis Bartolo.
\newblock {Emergent vortices in populations of colloidal rollers}.
\newblock {\em {NATURE COMMUNICATIONS}}, {6}, {JUN} {2015}.

\bibitem{Belovs:2014}
M.~Belovs and A.~Cebers.
\newblock {Relaxation of polar order in suspensions with Quincke effect}.
\newblock {\em {Phys. Rev. E}}, {89}({5}), {MAY 20} {2014}.

\bibitem{Yeo-Lushi-Vlahovska:2014}
K.~Yeo, E.~Lushi, and P.~M. Vlahovska.
\newblock Collective dynamics in a binary mixture of hydrodynamically coupled
  microrotors.
\newblock {\em Phys. Rev. Lett.}, 114:188301, 2015.

\bibitem{Cebers:2004}
A.~C\ifmmode~\bar{e}\else \={e}\fi{}bers.
\newblock Bistability and ``negative'' viscosity for a suspension of insulating
  particles in an electric field.
\newblock {\em Phys. Rev. Lett.}, 92(3):034501, Jan 2004.

\bibitem{Lemaire:2008}
E.~Lemaire, L.~Lobry, and N.~and Pannacci.
\newblock Viscosity of an electro-rheological suspension with internal
  rotations.
\newblock {\em J. Rheology}, 52:769--783, 2008.

\bibitem{Huang-Zahn-Lemaire:2011}
H-F. Huang, M.~Zahn, and E.~Lemaire.
\newblock Negative electrorheological responses of micro-polar fluids in the
  finite spin viscosity small spin velocity limit. i. couette flow geometries.
\newblock {\em J. Electrostatics}, 69:442--455, 2011.

\bibitem{Lemaire:2009b}
N.~Pannacci, E.~Lemaire, and L.~Lobry.
\newblock Dc conductivity of a suspension of insulating particles with internal
  rotation.
\newblock {\em Eur. Phys. J. E}, 28:411--417, 2009.

\bibitem{Melcher-Taylor:1969}
J.~R. Melcher and G.~I. Taylor.
\newblock Electrohydrodynamics - a review of role of interfacial shear stress.
\newblock {\em Annu. Rev. Fluid Mech.}, 1:111--146, 1969.

\bibitem{Cebers:2000}
A.~Cebers, E.~Lemaire, and L.~Lobry.
\newblock Electrohydrodynamic instabilities and orientation of dielectric
  ellipsoids in low-conducting fluids.
\newblock {\em Phys. Rev. E}, 63:016301, 2000.

\bibitem{Jones:1984}
T.~B. Jones.
\newblock Quincke rotation of spheres.
\newblock {\em IEEE Trans. Industry Appl.}, 20:845--849, 1984.

\bibitem{Turcu:1987}
I.~Turcu.
\newblock Electric field induced rotation of spheres.
\newblock {\em J. Phys. A: Math. Gen.}, 20:3301--3307, 1987.

\bibitem{Cebers:2002}
A.~Cebers.
\newblock Dynamics of an elongated magnetic droplet in a rotating field.
\newblock {\em Phys. Rev. E}, 66(6):061402, Dec 2002.

\bibitem{Dolinsky-Elperin:2005}
Y.~Dolinsky and T.~Elperin.
\newblock Equilibrium orientation of an ellipsoidal particle inside a
  dielectric medium with a finite electric conductivity in the external
  electric field.
\newblock {\em Phys. Rev. E}, 71:056611, 2005.

\bibitem{Dolinsky-Elperin:2006}
Y.~Dolinsky and T.~Elperin.
\newblock Dynamics of a spheroidal particle in a leaky dielectric medium in an
  ac electric field.
\newblock {\em Phys. Rev. E}, 73:066607, 2006.

\bibitem{Dolinsky-Elperin:2009b}
Y.~Dolinsky and T.~Elperin.
\newblock Stability of particle rotation in a rotating electric field.
\newblock {\em Phys. Rev. E}, 79:026602, 2009.

\bibitem{Dolinsky-Elperin:2009}
Y.~Dolinsky and T.~Elperin.
\newblock Electrorotation of a leaky dielectric spheroid immersed in a viscous
  fluid.
\newblock {\em Phys. Rev. E}, 80:066607, 2009.

\bibitem{JonesTB}
T.~B. Jones.
\newblock {\em Electromechanics of particles}.
\newblock Cambridge University Press, New York, 1995.

\bibitem{Adamthesis}
A.~Musial.
\newblock Nonlinear electrohydrodynamics of non-spherical particles.
\newblock {\em MS Thesis}, Dartmouth College, 2011.

\bibitem{Cebers:1991}
A.~Tsebers.
\newblock Chaotic solutions for the relaxation equations of electrical
  polarization.
\newblock {\em Magnetohydrodynamics}, 27:251--258, 1991.

\bibitem{Cebers:1993}
A.~Cebers.
\newblock Chaos in polarization relaxation of a low-conducting suspension of
  anisotropic particles.
\newblock {\em J. Magn. Magn. Mat.}, 122:277--280, 1993.

\end{thebibliography}

\end{document}